\documentclass{aa}
\usepackage{natbib}
\usepackage{amssymb}
\usepackage{amsmath}
\usepackage{graphicx}
\def\aap{A\& A}

\def\nat{Nature}
\def\apj{ApJ}

\def\max{\mathrm{max}}

\def\gram{\hbox{g}}
\def\cm{\hbox{cm}}

\def\kel{\hbox{K}}

\def\AU{\hbox{AU}}

\def\in{\mathrm{in}}
\def\out{\mathrm{out}}
\def\eff{\mathrm{eff}}
\def\LI{LI}
\def\ALI{ALI}
\def\VEF{VEF}
\def\EAMO{MEMO}
\def\SED{SED}
\def\comma{\,,}
\def\fullstop{\,.}
\def\thttle{Vertical structure models of T Tauri and Herbig Ae/Be disks}
\begin{document}
\title{\thttle}
\author{C.P.~Dullemond, G.J.~van Zadelhoff and A.~Natta}
\authorrunning{Dullemond, Van Zadelhoff \& Natta}
\titlerunning{\thttle} 
\institute{Max Planck Institut f\"ur Astrophysik, P.O.~Box 1317, D--85741 
Garching, Germany; e--mail: dullemon@mpa-garching.mpg.de \and 
Leiden Observatory, P.O.~Box 9513, 2300 Leiden, The Netherlands; e--mail:
zadelhof@strw.leidenuniv.nl \and Osservatorio Astrofisico di Arcetri,
Largo E.~Fermi 5, 50125 Firenze, Italy}
\date{DRAFT, \today}

\abstract{In this paper we present detailed models of the vertical structure
(temperature and density) of passive irradiated circumstellar disks around T
Tauri and Herbig Ae/Be stars. In contrast to earlier work, we use full
frequency- and angle-dependent radiative transfer instead of the usual
moment equations.  We find that this improvement of the radiative transfer
has strong influence on the resulting vertical structure of the disk, with
differences in temperature as large as 70\%.  However, the spectral energy
distribution (SED) is only mildly affected by this change. In fact, the SED
compares reasonably well with that of improved versions of the Chiang \&
Goldreich (CG) model. This shows that the latter is a reasonable model for
the SED, in spite of its simplicity. It also shows that from the SED alone,
little can be learned about the vertical structure of a passive
circumstellar disk. The molecular line emission from these disks is more
sensitive to the vertical temperature and density structure, and we show as
an example how the intensity and profiles of various CO lines depend on the
adopted disk model.  The models presented in this paper can also serve as
the basis of theoretical studies of e.g.~dust coagulation and settling in
disks.}

\maketitle


\section{Introduction}
The dust continuum emission often observed from T Tauri stars and Herbig
Ae/Be stars is widely believed to originate from a dust/gas disk surrounding
these stars (see e.g.~Beckwith \& Sargent,\citeyear{beckwithsargent:1996}). 
These disks presumably have  mass 
between $10^{-4}M_{\odot}$ and few$\times 10^{-1}M_{\odot}$, and are thought to
be the disks of dust and gas from which the stars were formed. At early
stages of the disk's evolution, the disks may still accrete, and lose mass
to the central star (Calvet, Hartmann, \& Strom \cite{calvethartstrom:2000}). 
The energy released during this process is
then the main source of power for the dust continuum emission from the
disk. At later stages of the disk's evolution the mass accretion rate drops,
and the outer regions of the disk start to become dominated by stellar
irradiation instead of viscous energy release. As the accretion rate drops
even further, more and more of the disk becomes ``passive'', and irradiation
eventually will be the only source of heating for the disk. As was discussed
by Kenyon \& Hartmann (\citeyear{kenyonhart:1987}), such a passive disk can
adopt a flaring shape, i.e.~a shape in which the ratio of the vertical
surface height $H_s(R)$ over $R$ increases with $R$, the distance from the
central star. In
this way the disk's surface always ``sees" the stellar surface,
and therefore intercepts some fraction of the stellar radiation at all radii.

The detailed vertical structure of such a disk can be computed by
solving the equations of vertical pressure balance coupled to the
equations of radiative transfer. Such computations have been done by
e.g.~D'Alessio et
al.~(\citeyear{dalessiocanto:1998},\citeyear{dalessiocalvet:1999}),
and Bell et
al.~(\citeyear{bellcassklhen:1997},\citeyear{bell:1999}). Their models
include detailed physics, including viscous dissipation, heating by
cosmic rays, and a study of the effects of self-gravity. A basic
weakness of these models is their simplified treatment of radiative
transfer, which uses the frequency-integrated moment equations in the
Eddington approximation with Planck- and Rosseland-mean opacities. The
reason for this simplification is that solving the full angle- and
frequency-dependent radiative transfer equations is a challenging
technical problem.  It is known that straightforward methods for
radiative transfer (e.g.~Lambda Iteration and Monte Carlo methods)
converge extremely slowly at high optical depth, and one is never
completely sure that a true solution has been reached.

In the field of stellar atmospheres these problems are well known, and many
different sophisticated radiative transfer algorithms have been developed
over the years. Such algorithms include Accelerated Lambda Iteration,
Complete Linearization methods, and Variable Eddington Factor methods (for a
review of radiative transfer techniques, see e.g.~Kudritzki \& Hummer
\citeyear{kudrhummer:1990}; Hubeny \citeyear{hubeny:1992}; Mihalas 
\citeyear{mihal:1978}). 

In particular the latter method has already been used successfully in
application to accretion disk models. Hubeny (\citeyear{hubeny:1990}) showed
that when the Eddington factors and mean opacities are known, the disk
temperature at every optical depth can be expressed by a simple analytic
formula. He presented an analytical solution for the case in which the
Eddington approximation is adopted and the Planck- and Rosseland mean
opacities are used.  Malbet \& Bertout (\citeyear{malbetbertout:1991}) and
Malbet, Lachaume \& Monin (\citeyear{malbetlachaume:2001}) also presented
disk equations based on the variable Eddington factor method, and present
solutions to the full angle-dependent transfer problem. Their solutions,
however, are based on the assumption of a grey opacity, and therefore fall
short of our aims.  A full frequency-angle dependent vertical structure
model has been presented by Sincell \& Krolik (\citeyear{sinkrol:1997}) for
X-ray irradiated accretion disks around active galactic nuclei.
Technically, that work comes close to the kind of calculations we wish to
make in this paper, but in addition to the different physics involved,
their calculations only solve the upper irradiated skin of the disk.

It is the goal of this paper to present complete vertical structure models
for irradiated passive flaring disks based on full angle-frequency-dependent
radiative transfer and vertical hydrostatic equilibrium. We use the
method of Variable Eddington Factors as our main radiative transfer
algorithm, and we present a variant of this algorithm that works fast and is
stable under all circumstances. An Accelerated Lambda Iteration (\ALI)
algorithm is used to check the results and make sure that a true solution to
the transfer equation is obtained.

\section{The disk model}
The structure equations for a passive irradiated disk can be grouped in
three sets. First we have the equations describing the transfer of primary
(stellar) photons as they move from the star radially outwards, and
eventually get absorbed by the dust grains in the upper layers of the
disk. The energy absorbed in these surface layers will be re-emitted at
infrared wavelengths. Half of this radiation will travel upwards and escape
from the disk. The other half will move downwards into the disk, where it
will once again be absorbed and re-emitted. Since the disk's optical depth
can be rather high, in particular at short wavelengths, this process can
repeat itself many times.  In this way the energy diffuses all the way down
to the equatorial plane of the disk. This process of radiative diffusion can
be well described by the equations for plane-parallel 1-D vertical radiative
transfer. It is here that most treatments in the literature use the moment
equations with the Eddington approximation and mean opacities (henceforth
\EAMO{} method). In this paper we treat this problem in a fully
angle-frequency dependent way. Once this problem has been solved and the
temperature stratification has been found, we can proceed to solve the third
and final equation: the equation of vertical pressure balance.

These three coupled sets of equations are solved iteratively: we move from
stage 1 (primary stellar radiation), to stage 2 (diffuse radiation field) to
stage 3 (vertical pressure balance), and back to stage 1. This is repeated
until the relative difference in the density between successive iterations drops below
the convergence criterion, generally taken to be $10^{-2}$.

\subsection{Stage 1: Extinction of primary photons}
The impinging of primary (stellar) radiation onto the surface of the disk is
modeled using a ``grazing angle'' recipe, similar to that used by
D'Alessio et al.~(\citeyear{dalessiocanto:1998}) and Chiang \& Goldreich
(\citeyear{chianggold:1997}, henceforth CG97). Deliberately we do not use 
full 2-D/3-D ray-tracing, since this procedure can lead to numerical 
instabilities (Dullemond in prep.). The
``grazing angle'' recipe models the irradiation using vertical
plane-parallel radiative transfer, with the radiation entering the disk
under an angle $\beta(R)$ with respect to the disk's surface. 
At each radius $R$ and each vertical
height $z$ we evaluate the local primary stellar flux:
\begin{equation}
F_\nu(R,z) = \frac{L_\nu}{4\pi R^2} \exp\left(-\tau_\nu(R,z)/\beta(R)\right)
\comma
\end{equation}
where $L_{\nu}$ is the stellar luminosity and the vertical optical depth
$\tau_\nu(R,z)$ is defined as
\begin{equation}
\tau_\nu(R,z)= \int_z^\infty \rho(R,z)\kappa_\nu dz
\comma
\end{equation}
with $\rho(R,z)$ the dust density (g cm$^{-3}$) and $\kappa_{\nu}$ the dust
absorption opacity (cm$^{2}$ g$^{-1}$). We ignore scattering, and we ignore
the higher order geometrical effects which start to play a role when
$z/R\gtrsim 1$. The grazing angle $\beta(R)$ is defined as $\beta(R)=
0.4\,R_{*}/R + R\,d(H_s/R)/dR$, where $H_s$ is the surface height of the
disk and $R_{*}$ the stellar radius (see CG97).  We define $H_s$ as the
height above the midplane where 63\% (i.e., $1-\exp(-1)$) of the integrated
stellar radiation has been absorbed.  Although this definition is somewhat
arbitrary, it is not critical for the results presented in this paper.

It is convenient to express $\beta(R)$ as
\begin{equation}\label{eq-def-betaangle}
\beta(R) = 0.4\, \frac{R_{*}}{R}+\xi(R)\frac{H_s}{R}
\comma
\end{equation}
where $\xi$ is the flaring index defined as:
\begin{equation}
\xi\equiv\frac{d\lg(H_s/R)}{\lg(R)}
\fullstop
\end{equation} 
This flaring index is a number of order $\xi\simeq 2/7$ (CG97). Its value is
computed self-consistently during the iteration procedure. Usually the
iteration is started with a guess for $\xi$ (for instance $\xi=2/7$), and
updated after each few iteration steps. In order to avoid numerical
instabilities, $\xi$ is always evaluated two radial gridpoints away from the
point where it is used (see appendix of Chiang et
al.~\citeyear{chiangjoung:2001}).

It is important to note that a proper self-consistent computation of the
flaring index $\xi$ is crucial if one wants to achieve energy conservation.
Assuming it to be some fixed value is guaranteed to result in disks
that emit more (or less) radiation than they receive.

\subsection{Stage 2: Vertical radiative transfer}
Once the function $F_\nu(R,z)$ is known, one can compute the amount of
absorbed primary radiation per volume element:
\begin{equation}
q(R,z)=\int_0^\infty \rho(R,z) \kappa_\nu F_\nu(R,z) d\nu
\end{equation}
This energy will then be re-emitted as infrared radiation, half of which
will diffuse towards lower $z$ into the disk interior. The transfer of this
re-emitted IR radiation through the disk can be approximated as a 1-D slab
geometry transfer problem of high optical depth.

The intensity of this diffuse infrared radiation field is denoted by
$I_{\mu,\nu}$ and obeys the following radiative transfer equation:
\begin{equation}\label{eq-trans-de}
\mu\frac{dI_{\mu,\nu}}{dz} = \rho \kappa_\nu ( B_\nu(T) - I_{\mu,\nu})
\comma
\end{equation}
where $\kappa_\nu$ is the opacity per unit of matter, $\rho$ is the material
density and $B_\nu(T)$ is the Planck function. This differential equation
has to be integrated along $z$ for all $\nu$ and $\mu$, in order to obtain
the  intensity function $I_{\mu,\nu}(z)$. The dust temperature $T$ is
determined by assuming thermodynamic equilibrium with the radiation field:
\begin{equation}\label{eq-therm-equil}
\int_0^\infty \rho\kappa_\nu B_\nu(T) d\nu = 
\int_0^\infty \rho\kappa_\nu J_\nu d\nu + \frac{q}{4\pi}
\comma
\end{equation}
where the mean intensity $J_\nu$ is defined as
\begin{equation}\label{eq-mean-int}
J_\nu(z) = \frac{1}{2}\int_{-1}^{1} I_{\mu,\nu}(z) d\mu
\fullstop
\end{equation}

Equations (\ref{eq-trans-de},\ref{eq-therm-equil},\ref{eq-mean-int}) form a
set of coupled integro-differential equations in the $\mu$, $\nu$, $z$
space. A straightforward way to solve them would be to use the method of
Lambda Iteration (\LI): first evaluate $I_{\mu,\nu}(z)$
(Eq.~\ref{eq-trans-de}), then $J_\nu(z)$ (Eq.~\ref{eq-mean-int}) and finally
$T(z)$ (Eq.~\ref{eq-therm-equil}), and iterate this procedure until
convergence is reached. However, it is well known that this method converges
very slowly at high optical depths (see e.g.~Rybicki \& Hummer
\citeyear{rybhum:1991}).

Instead, we use the method of variable Eddington factors (\VEF, see appendix
\ref{sec-vet-method} for a description of the method). This method estimates
the shape of the radiation field by using a moment equation, but eventually
reaches a solution that solves the full frequency- and angle-dependent
transfer problem. It converges generally within $\sim$10 iterations, and is
therefore much faster and more reliable than the \ALI{} method.

\subsection{Stage 3: vertical hydrostatic equilibrium}
Once the temperature at all locations in the disk is known, one can integrate
the pressure balance equation to find the density structure. The vertical
pressure balance equation is:
\begin{equation}
\frac{dP(R,z)}{dz} = -\rho(R,z)\frac{GM}{R^3}z
\fullstop
\end{equation}
Since $P=k\rho T/\mu m_u$ (with $\mu=2.3$ for a H$_{2}$, He mixture), and
the derivatives of $T$ are known, this equation can be cast into the form of
an integral in $\rho$. This integration is started at $z=0$, taking
$\rho(R,0)$ at first as an arbitrary value. One can then compute the
resulting vertical surface density:
\begin{equation}
\Sigma = 2 \int_0^\infty \rho(R,z) dz
\comma
\end{equation}
and then renormalize the  density profile to the actual surface density.
This new density structure is then compared to the 
density structure of the previous iteration.  If the convergence criterion
is not met, the iteration proceeds by going back to Stage 1.
This iterative procedure has proven to be stable and to converge quickly, typically after
about five to eight iterations.

\section{Resulting vertical structure}
As our example case we take a star with $T_{\eff}=3000\kel$,
$R_{*}=2.0\,R_{\odot}$ and $M_{*}=0.5\,M_{\odot}$. Our disk has an inner
radius of $R_{\in}=3\,R_{*}$, an outer radius of $R_{\out}=300\,\AU$ and a
gas+dust surface density $\Sigma=\Sigma_0\,(R/\AU)^{-1}$ with $\Sigma_0$ the
surface density at $1\,\AU$.
We assume that the gas and the dust are always mixed in the mass
ratio $100:1$, and we take for the dust opacity the opacity of astronomical
silicate of Draine \& Lee (\citeyear{drainelee:1984}) for grains of 
0.1 $\mu$m size.

\begin{figure*}
\centerline{
\includegraphics[width=9cm]{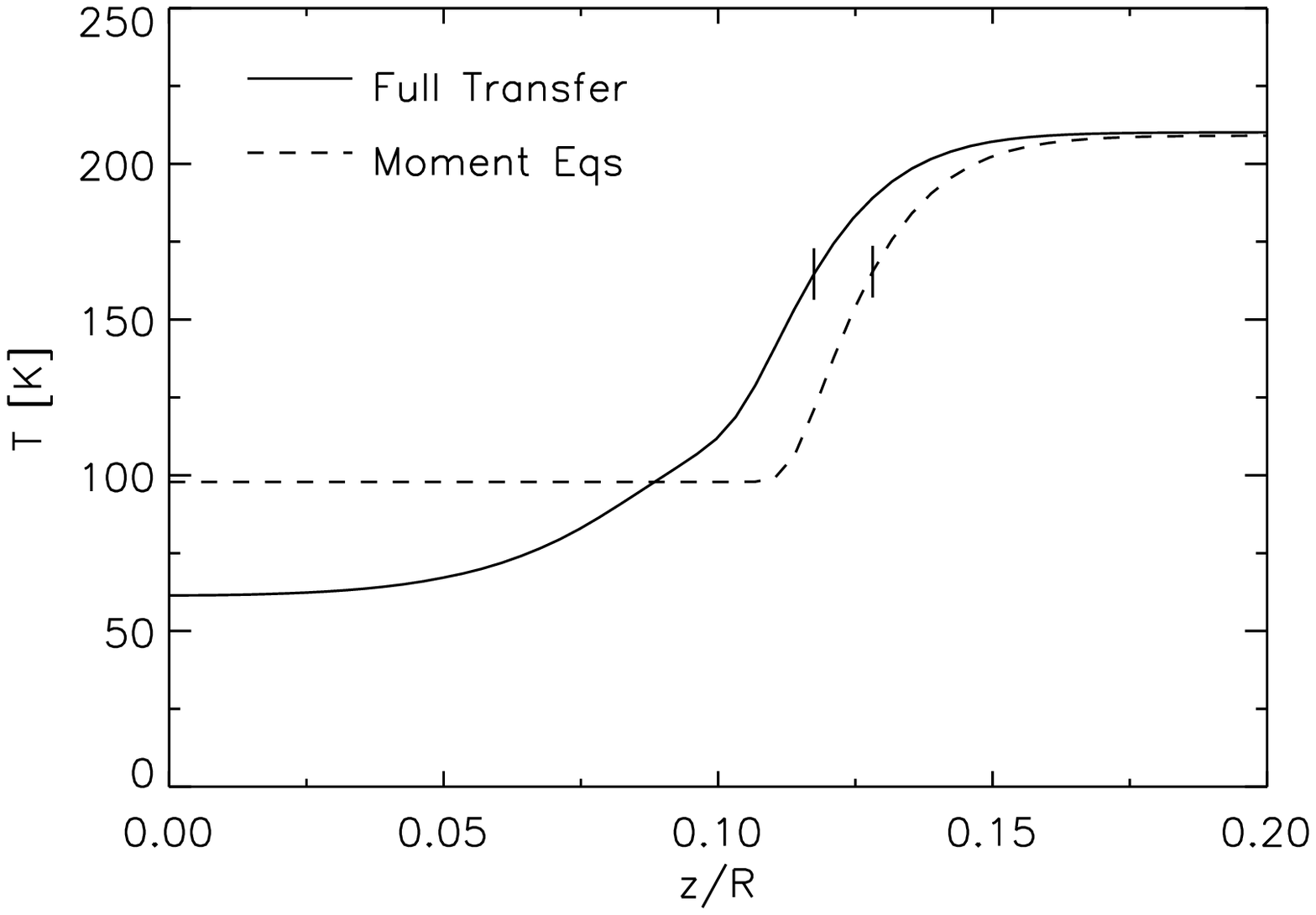}
\includegraphics[width=9cm]{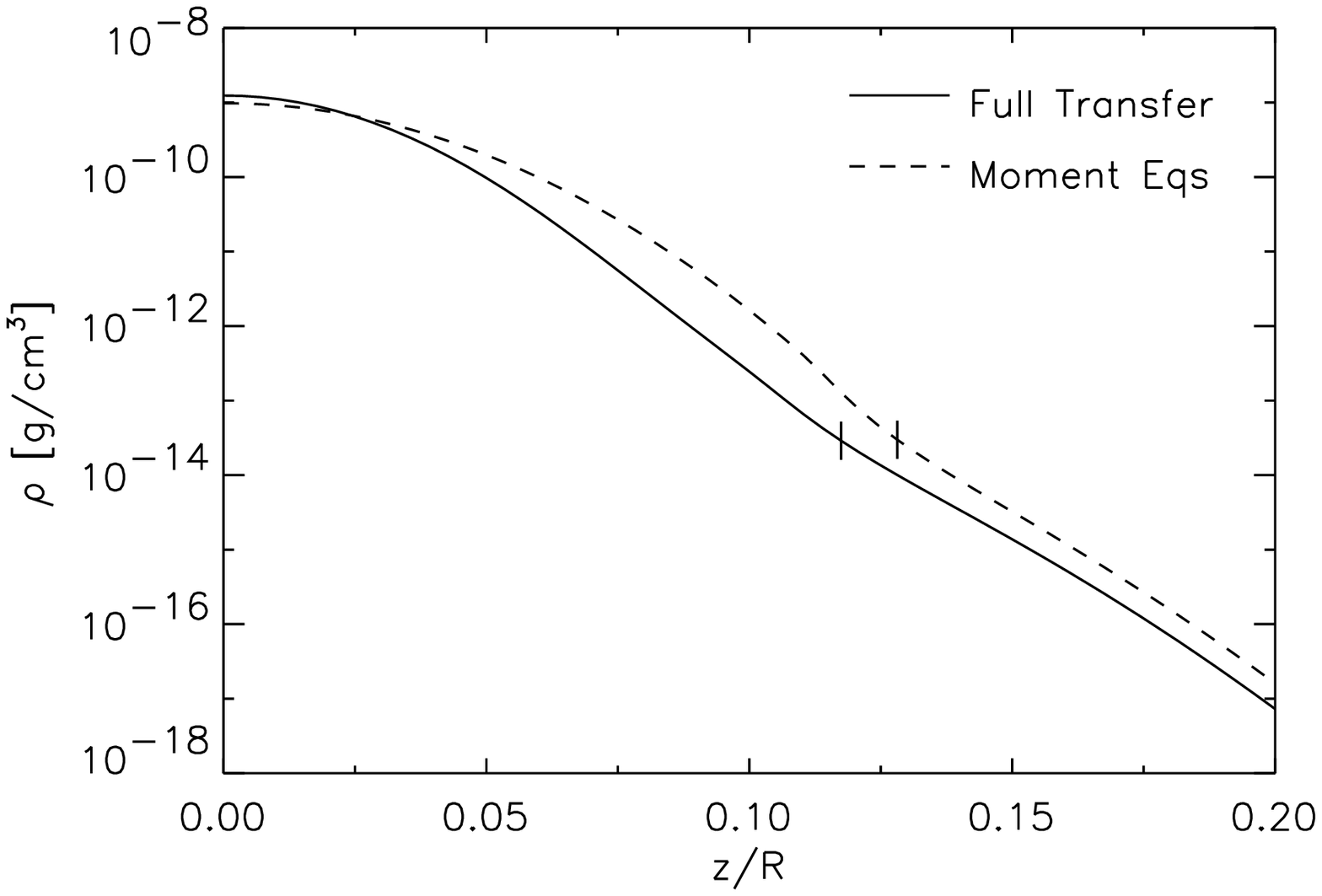}}
\caption{\label{fig-full-struct-slice} The vertical structure of a passive
irradiated flaring disk surrounding a star with $T_{\eff}=3000\kel$,
$R_{*}=2R_{\odot}$ and $M_{*}=0.5M_{\odot}$. The distance from the star at
which the structure is shown is 1 AU. The gas+dust surface density at this
radius is $\Sigma(1\AU)=10^3\gram/\cm^2$. The dashed line shows the
structure as computed using the moment method with mean opacities (the
\EAMO{} equations). The solid line shows the structure computed using full
angle-frequency dependent radiative transfer (using the \VEF{} method).
Left panel shows the temperature, while the right panel shows the density.
The tickmarks on the curves indicate the location of the defined disk
surface, i.e.~the $z/R$ above which 63\% of the direct stellar light has
been absorbed.}
\end{figure*}

In Fig.~\ref{fig-full-struct-slice} the vertical structure of the disk at a
distance of $1\AU$ from the star is shown. For this calculation we took
$\Sigma_0=10^3\,\gram/\cm^2$, which corresponds to a visual
optical depth through the disk $\tau_V=2.3\times 10^4$. 
To better illustrate the results, we compare then to those obtained
for the same parameters using the more approximated \EAMO{} method
of solution. The left panel shows the vertical temperature profile.
One sees that the full transfer model has a 
much smoother  structure than the \EAMO{} solution, and that  the
midplane temperature is significantly lower, at least for the high optical depth
case shown here. The main effect of the lower midplane
temperature is a slightly smaller pressure scale height at the equator 
($H_p/R=0.022$ for the VEF models versus $H_p/R=0.028$ for the \EAMO{} one,
as expected from the factor 1.57 difference in the midplane temperature
between the two models), so that
the disk is a bit more compressed toward the equatorial plane.  

The right panel of Fig.~\ref{fig-full-struct-slice}
shows the run of the density as function of $z/R$. One can see how the
two solutions differ only by a small factor near the midplane, since
the pressure scale heights are not very different. At higher
$z/R$ the differences become much larger (an order of magnitude at
$z/R \sim 0.075-0.1$). This can be understood as a slight vertical 
shift in $z$ of the steep density profile. 
\begin{figure*}
\centerline{
\includegraphics[width=9cm]{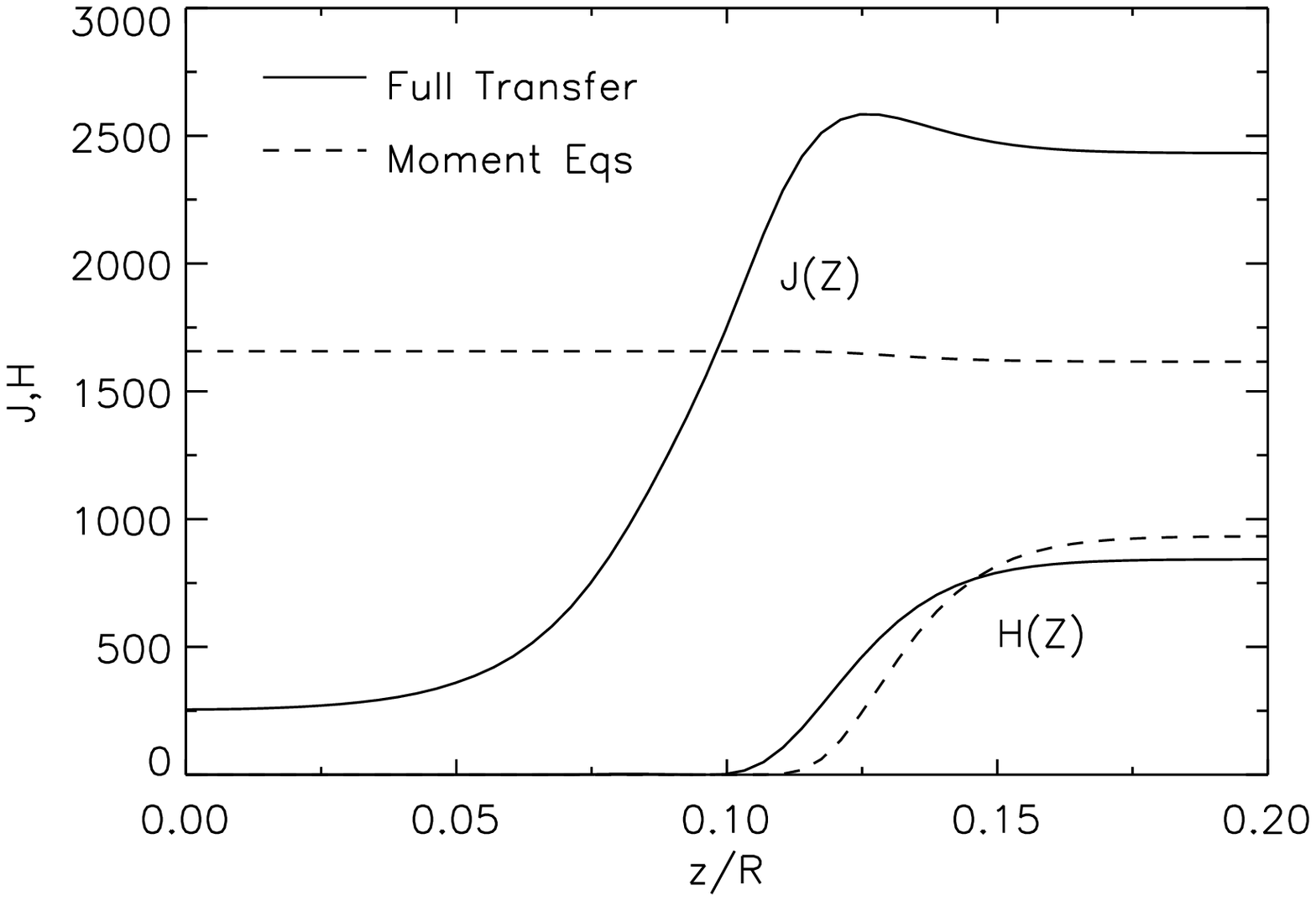}
\includegraphics[width=9cm]{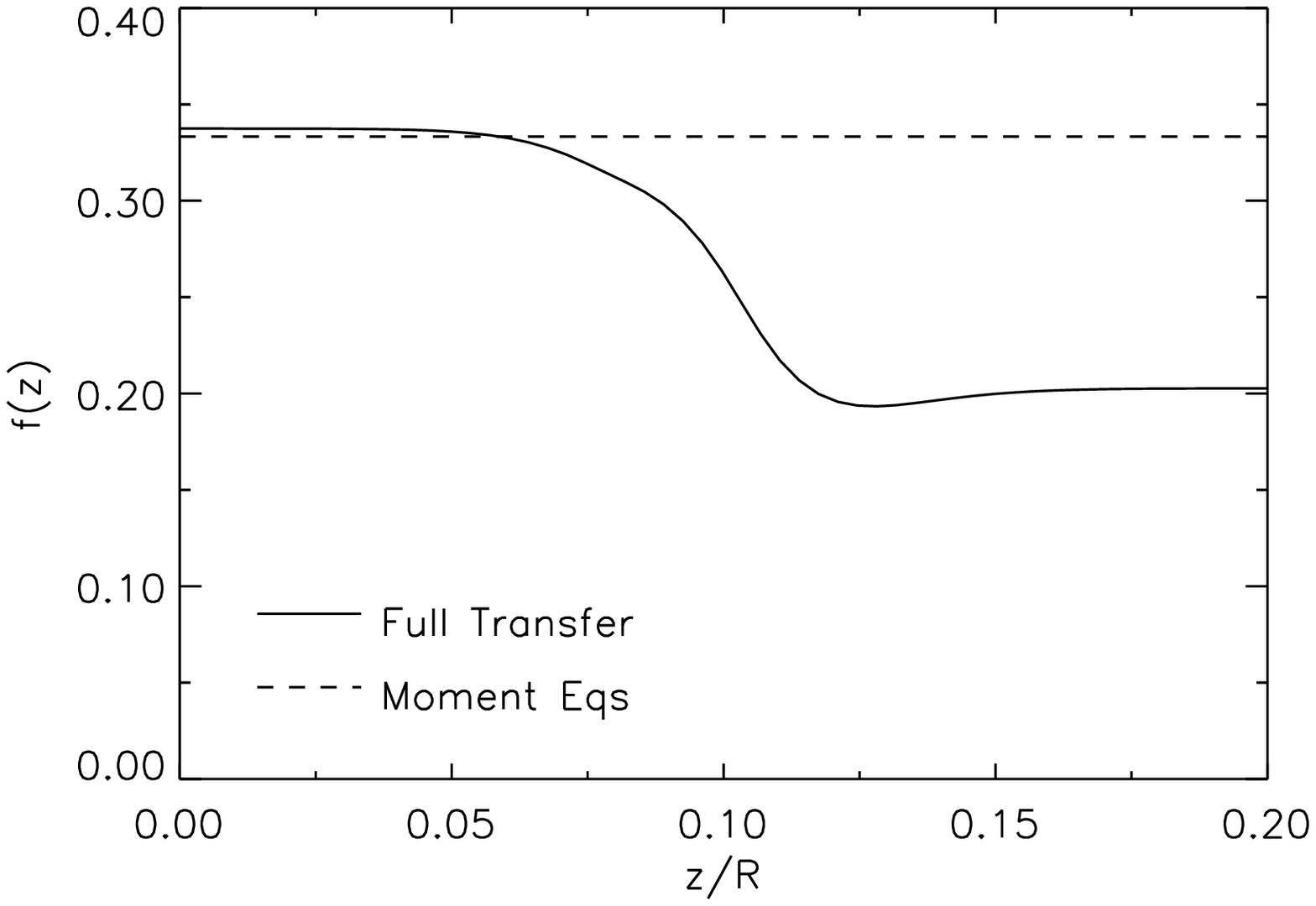}}
\caption{\label{fig-jh-eddfact-slice} The radiation field in the disk, as
computed using the moment equations (\EAMO{} method: dashed line) and the
full radiative transfer (\VEF{} method: solid line). The model and the
location of the slice (1 AU) are the same as in
Fig.\ref{fig-full-struct-slice}. Left panel the frequency-integrated mean
intensity $J$ and Eddington flux $H$. Right panel the frequency-averaged
Eddington factor
$f$. In the \EAMO{} method this factor is fixed by definition to $1/3$,
while in the \VEF{} method it is computed self-consistently. Note that the
slight deviation of $f$ from $1/3$ close to the equatorial plane is due to
the discretization of the radiation field in $\mu$, so that the integration
of $\mu^2$ over $\mu$ is not exactly $1/3$.}
\end{figure*}

Fig.~\ref{fig-jh-eddfact-slice}  illustrates some technical aspects of the
radiation transfer solutions, which clarify the results
just shown. Firstly, one can see (right panel)  how
the Eddington factor $f(z)$ drops below $1/3$ in the upper layers of the
disk. This is because in these optically thin layers, most of the intensity
is more or less parallel to the disk instead of pointing out of the
disk. This is an effect that is opposite to what happens for instance in
stellar winds, where most of the radiation eventually gets beamed outwards,
and the value of $f$ becomes close to $1$.

It is also clear from Fig.~\ref{fig-jh-eddfact-slice} (left panel)
that as a result of the use of the Rosseland
mean opacity in the \EAMO{} equations, the mean intensity in the \EAMO{}
method remains virtually constant, whereas in reality the mean intensity
drops strongly towards the equatorial plane. This results in a
non-isothermal disk structure, as opposed to what would be expected on the
basis of the usual diffusion approximation considerations. According to
these considerations one would expect to have a perfectly isothermal disk
interior since there are no sources of heat in the disk. Indeed, if a grey
opacity is used in the full transfer model as a test, the disk interior
becomes isothermal as expected. But as soon as non-grey dust opacities are
included, the isothermality is broken. The reason is that some radiation
is able to leak out of the disk at long wavelengths, where the opacity is
much smaller than at the wavelength of the bulk of the radiation. The
slight positive temperature gradient causes a downward diffusive flux at
shorter wavelengths that exactly cancels this energy leakage, so that
the total flux remains zero deep within the disk.

\begin{figure}
\centerline{
\includegraphics[width=9cm]{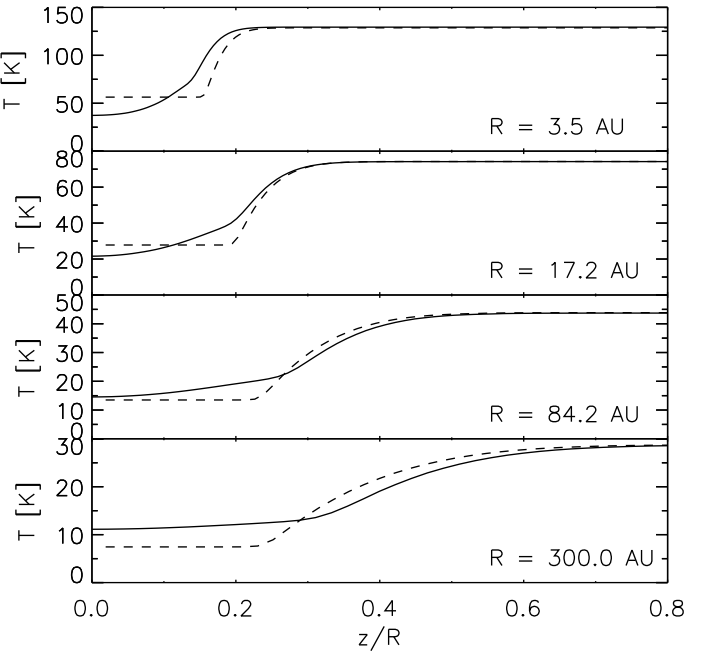}}
\caption{\label{fig-temp-all-radii} The vertical structure at different
radii, as computed using the moment equations (dashed line) and using full
angle-frequency dependent radiative transfer using the \VEF{} method (solid
line). On the horizontal axis the dimensionless vertical height.  The model
parameters are the same as in Fig.\ref{fig-full-struct-slice}}
\end{figure}

In Fig.\ref{fig-temp-all-radii} the vertical temperature structure is shown
for different values of the radius.  Near the star, where the disk optical
depth is very large, the equatorial plane temperature is usually below that
of the \EAMO{} model, whereas at large radii, where the disk becomes
optically thin to its own diffuse radiation, the \EAMO{} models predict
lower temperature values than what we find. The latter effect is related to
the fact that the \EAMO{} method treats the diffuse radiation field in a
frequency-integrated way, and has therefore no information on how the energy
is distributed over frequency. In other words: the \EAMO{} method uses
mean opacities based on the local dust temperature, whereas the local
radiation temperature may be different in the optically thin case.

In Fig.\ref{fig-temp-all-radii} it can also be seen that at small radii the
model-predicted  temperature for large $z/R$ is exactly constant. In reality 
we expect a small decrease of $T$
for increasing $z/R$ since the distance to the star
increases as $\sqrt{R^2+z^2}$. But since higher order geometrical effects
are ignored in our model this is not taken into account. Fortunately the
gas+dust density at large $z/R$ is very small, and this effect has no 
practical influence on the resulting spectra.

\begin{figure*}
\centerline{
\includegraphics[width=9cm]{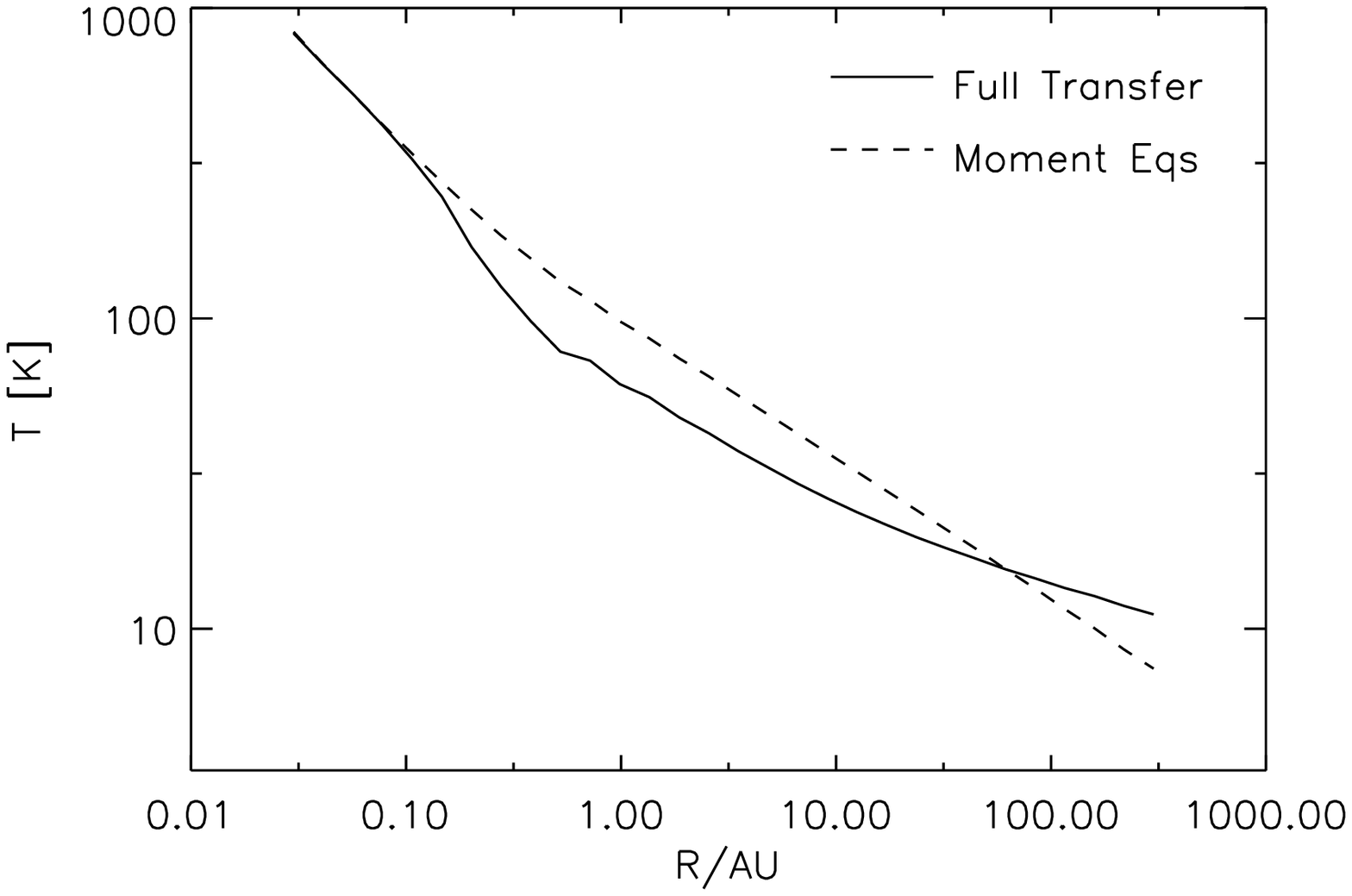}
\includegraphics[width=9cm]{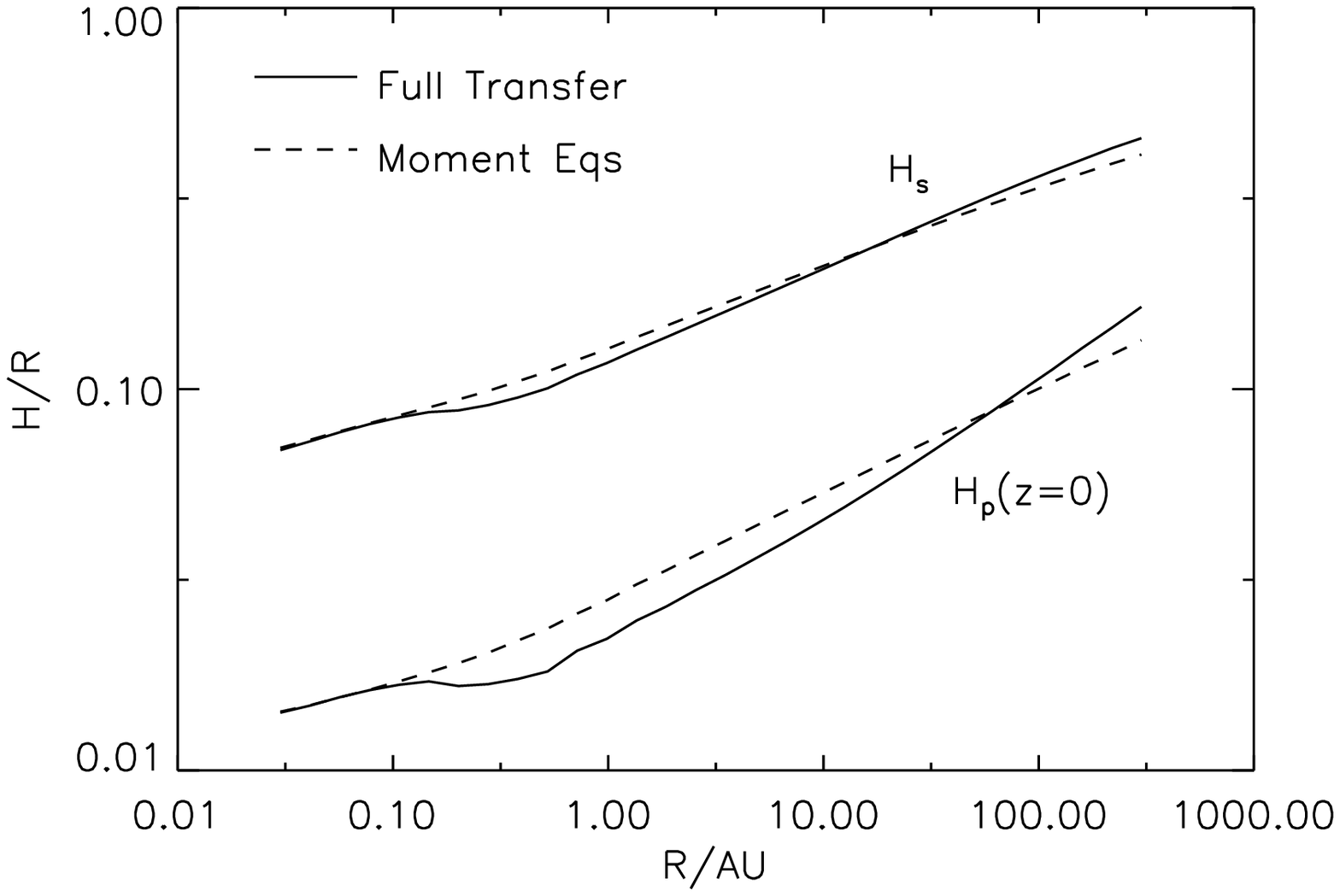}}
\caption{\label{fig-temp-equ-surfheight} The temperature at the equatorial plane
and the surface height of the disk, for the same model parameters as in
Fig.\ref{fig-full-struct-slice}.}
\end{figure*}

Finally, we show
in Fig.\ref{fig-temp-equ-surfheight} the radial dependence of
the equatorial plane temperature (left panel)
and of the pressure scale and surface height of the disk
for the VEF and \EAMO{} solutions (right panel).
As was already known from
Fig.\ref{fig-temp-all-radii}, the midplane temperature is lower
than in the \EAMO{} approximation at most radial intervals, but
becomes higher at the outer edge. However, at the very inner
edge it is again the same in both approaches. This is because at these
radii the vertical optical depth of the disk becomes so large that
even at millimeter wavelengths there is no possibility for the gas to
cool, and the disk becomes isothermal.
Fig.\ref{fig-temp-equ-surfheight}, right panel, shows the radial
profile of the pressure scale height, which follows approximately
that of the midplane temperature, as discussed, and of
the surface height. It is interesting to note how the
\EAMO{} approximation and the
full transfer model give  almost equal values of $H_s$. 
This is a result of the higher temperatures at higher elevations above the
midplane. This can be seen from the fact that the upper part of the
temperature rise, which is where the direct stellar radiation is absorbed, 
is located at about the same $z$ in both models. 

\begin{figure}
\centerline{
\includegraphics[width=9cm]{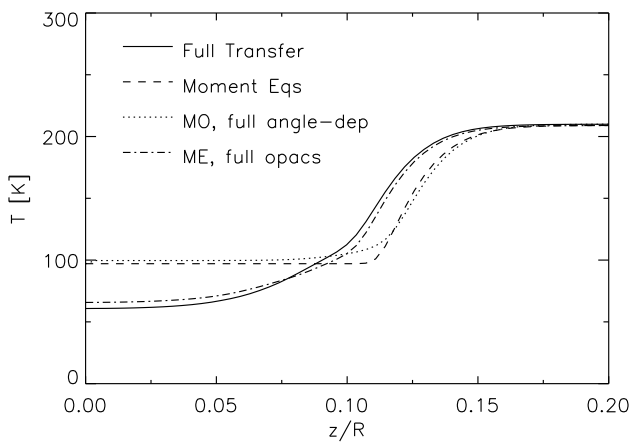}}
\caption{\label{fig-alt-methods} Same figure as
Fig.\ref{fig-full-struct-slice}, but now compared to two additional
methods. The dotted line is for the method using full $\mu$-dependence, but
the usual mean opacities (Planck and Rosseland). The dot-dashed line is for
the method using the full frequency-dependent opacities (and therefore
frequency-dependent transfer), but using the Eddington approximation for the
angular dependence.}
\end{figure}

With these differences between the simple (\EAMO) and the complex (VEF)
method, one may wonder whether a compromise between the two methods may give
results that are close enough to the real one. Even with the efficiency of
the VEF method, solving the full angle-dependent radiative transfer problem
is a time-consuming process. It may therefore not be very practical for
e.g.~the computation of the evolution and/or the dynamics of the disk, where
the radiative transfer problem needs to be solved thousands of times. In
Fig.\ref{fig-alt-methods} we show the same slice as in
Fig.\ref{fig-full-struct-slice}. Here we added the computed temperature
profiles for two alternative methods: one in which the full $\mu$-dependence
is solved, but with Planck- and Rosseland mean opacity replacing the
$\kappa_J$ and $\kappa_H$, and one in which the Eddington approximation is
used, but with the full frequency-dependent opacities.

One sees that the former does not improve much on the \EAMO{} method. But
the latter in fact gets very close to the 'right' answer. This shows that
most of the problems with the \EAMO{} method originate from the wrong
treatment of mean opacities, while the use of the Eddington approximation
iself ($f_\nu=1/3$ and $H(z=z_{\max})=J(z=z_{\max})/\sqrt{3}$) does not
cause major problems.  We conclude that an approximate method, in which
frequency-dependent radiative transfer is done in the Eddington
approximation, is in fact a relatively accurate method.  This method avoids
the CPU-time consuming full frequency-angle dependent radiative transfer,
and is therefore faster than the full VEF method. Nevertheless, in this
paper we adopt the complete VEF method, since it is still fast enough for
our purpose, and it is exact, at least in the 1-D approximation made here.

\subsection{Effect on the spectral energy distribution}
The new vertical structure models have a spectral energy distribution (\SED)
that is not very different from the ones predicted by the vertical
structures from the \EAMO{} approximation. In fact, they do not even differ
much from the even simpler semi-analytic treatment of the Chiang \&
Goldreich model, if certain modifications are made to the latter (see Chiang
et al.~\citeyear{chiangjoung:2001} and Dullemond, Dominik \& Natta
\citeyear{duldomnat:2001}, henceforth DDN01).
\begin{figure}
\centerline{
\includegraphics[width=8cm]{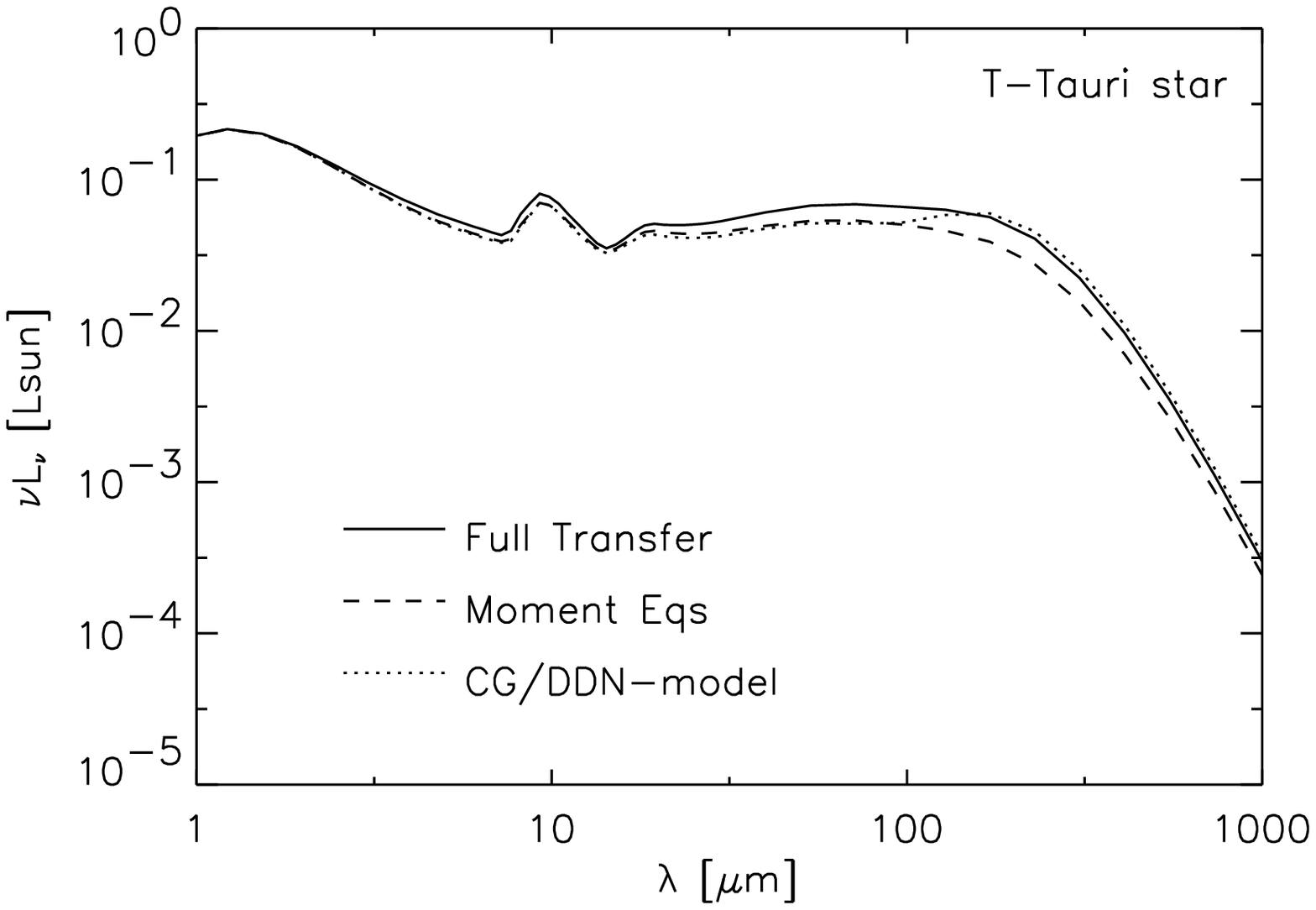}}
\centerline{
\includegraphics[width=8cm]{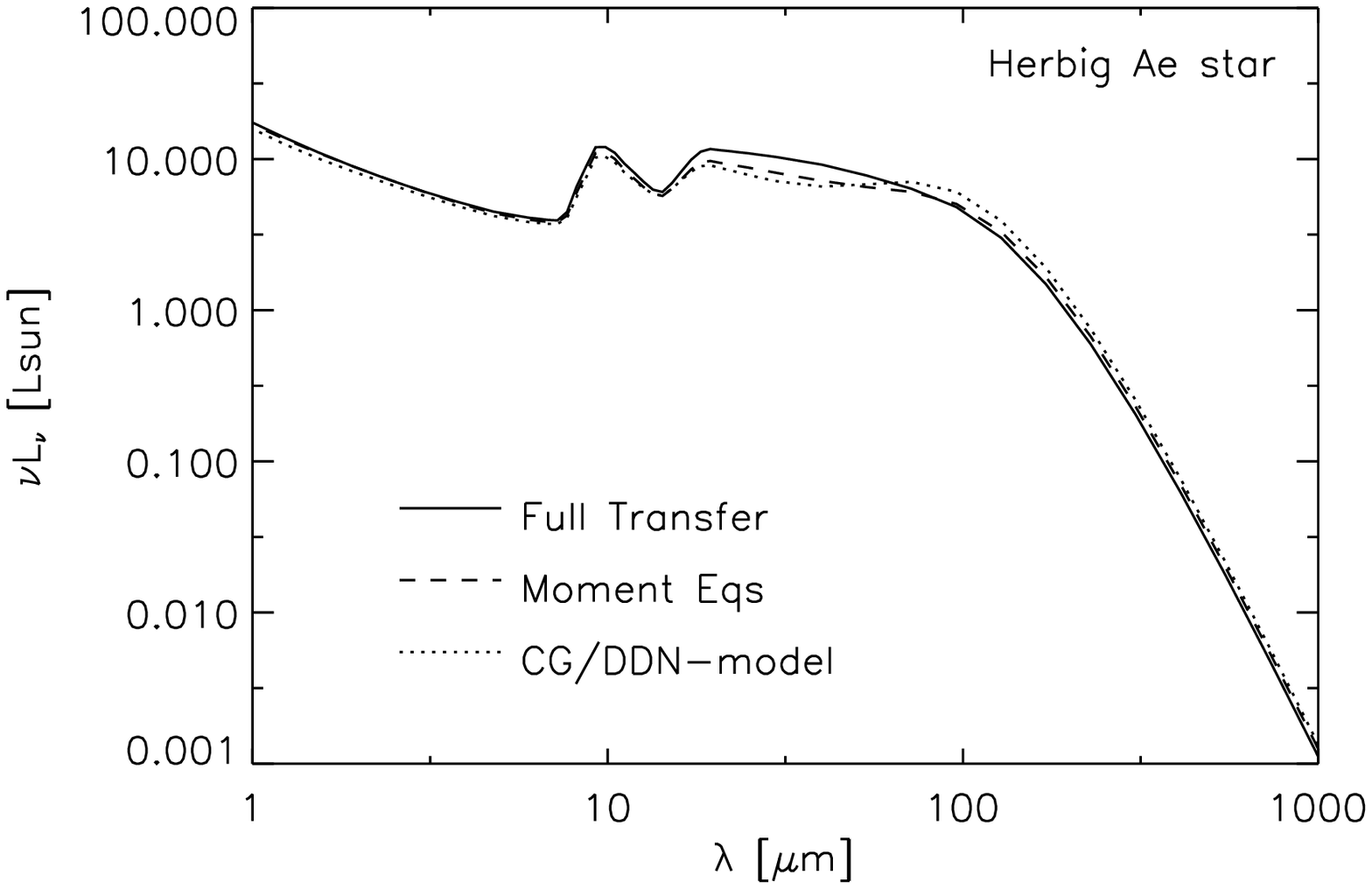}}
\caption{\label{fig-sed-compare} The spectral energy distribution computed
using the \EAMO{} method (dashed line), using the CG97/DDN01 model (dotted
line) and using full angle-frequency dependent radiative transfer with the
\VEF method (solid line). For all models we use a face-on viewing angle
($i=0$). Upper panel: SED for a T Tauri star with
$T_{\eff}=3000\kel$, $R_{*}=2R_{\odot}$ and $M_{*}=0.5M_{\odot}$. Lower
panel: SED for a Herbig Ae star with $T_{\eff}=9520\kel$,
$R_{*}=2.5R_{\odot}$ and $M_{*}=2.4M_{\odot}$. For both cases the disk has
$\Sigma=10^3(R/\AU)^{-1}$. Note that in the latter, we did not make a
special treatment for the inner edge of the disk, as is required (DDN01),
hence the absence of the near-IR bump.}
\end{figure}
This may be rather surprising, since the disk structure is so drastically
different, with equatorial temperatures differing up to 70\%.
The explanation of this lies in the fact that the total emitted flux of the
disk is prescribed by energy balance. At each radius it is expected that the
disk emits both an optically thin component and an optically thick one. Each
component carries half of the emitted flux, which equals the total absorbed
flux. This means for instance that the optically thick component (at far-IR
wavelengths) must have an {\em effective} temperature that is the same in
all models. In the full transfer models the effective temperature is
therefore the same as the effective temperature of the Chiang \& Goldreich
model, even though the spectrum may not be a perfect blackbody.

There are however some differences that are worth noting.  The full transfer
models predict higher fluxes in the mid and far-infrared than MEMO
models. The difference is of the order of 30\% in the wavelength range
between 50 and 500 $\mu$m for a T Tauri star, and between 50 and 100 $\mu$m
for a hotter Herbig Ae star.
However, the shape of the SED is not very different, even if the full
transfer models fall slightly more rapidly towards the far infrared.  The
SEDs of the CG97/DDN01 models are somewhat flatter in the same
range of wavelengths, since they are composed of two equally strong
components (see the case of the HAe star, where the two ``bumps" are clearly
visible).  This is a direct consequence of the discrete two-layered
structure of the CG97/DDN01 model. In the full transfer model the surface
layer has a continuous temperature gradient, which smoothly matches to the
photosphere. This produces the smoother SEDs at far-IR wavelengths. The
bumps in the SEDs of the CG97/DDN01 model are therefore an artifact of the
adopted two-layers simplification.

The region aroung the 10 micron feature are virtually not affected by the
full treatment of radiative transfer. The emission in this region is
dominated in all models by emission from the optically thin
surface layers, whose properties
are quite
independent of what happens deep below in the disk, and is therefore much
less affected by the different methods of solving the radiative transfer.

\subsection{Effects on molecular line intensities}
Knowledge of the correct vertical structure of disks (temperature and
density) is of great importance for predicting the abundances of
different molecular species and the intensity of their lines.  In a
separate paper we discuss the differences between various models in
detail, in the context of molecular line observations from T Tauri and
Herbig Ae stars, for which the \SED{} is sufficiently well known
(Van Zadelhoff \& Dullemond in prep). Here we confine ourselves to a
demonstration of the effects on the intensity of the lines of CO
and its two main isotopomers $^{13}$CO and C$^{18}$O, assuming that
gas and dust temperatures are the same. The
CO molecule can be used as a temperature tracer, and is therefore
suited to probe the differences between the two models.
\begin{figure}
\centerline{
\includegraphics[width=9cm]{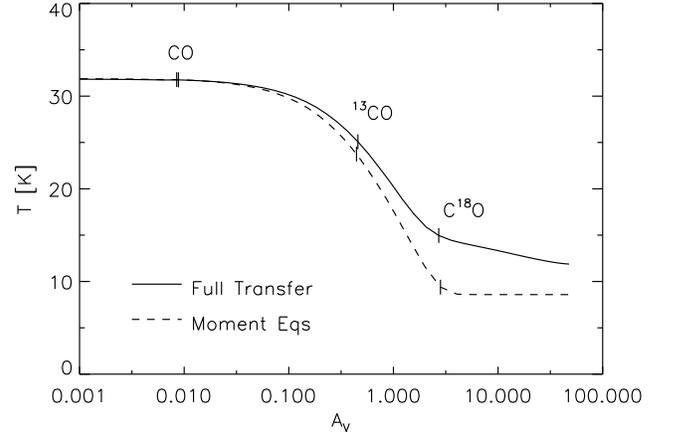}}
\caption{\label{fig-t-over-col} The temperature of the disk at a radius of
$220\,\AU$, as a function of vertical $A_V$ into the disk.  Here $A_V=0$
corresponds to $z=\infty$. The tickmarks show te $\tau=2/3$ locations for
the 3-2 line of CO, $^{13}$CO and C$^{18}$O respectively.}
\end{figure}

A first insight of how line intensities depend on the disk model can
be obtained from Fig.~\ref{fig-t-over-col}, which shows how the
temperature varies as a function of the gas column density (for our
template T Tauri star model and $R$=220 AU). If we assume for
simplicity that the line optical depth is proportional to the gas column
density, and that the brightness temperature in the line is roughly
equal to the gas temperature where the optical depth in the line is
$\tau\sim 2/3$, one can immediately see that different models will
predict the same brightness temperature for lines that are very
optically thick, but that MEMO models will underestimate significantly
the brightness temperature of more optically thin lines, which form
more deeply in the disk where the predicted temperatures differ. 
\begin{figure}[ht!]
\centerline{
\resizebox{\hsize}{!}{\includegraphics{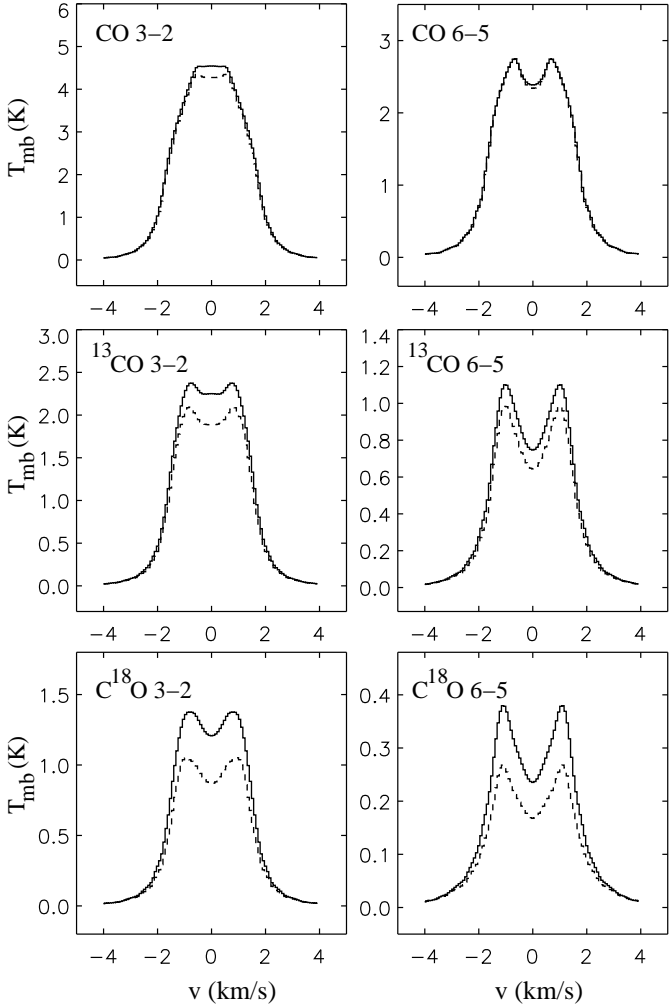}}}
\caption{\label{fig-line} The molecular line emission for the 3--2
(left) and 6--5 (right) transitions for the molecules CO, $^{13}$CO
and C$^{18}$O emitted by a disk at 150 pc. The disk is assumed to be
seen under an inclination of 60\degr, with a beam of 4.0\arcsec.  The
solid line represents the emission using the full continuum transfer,
the dashed line shows the results using the moments equations.}
\end{figure}

For a true comparison of models to the observations a number of effects have
to be taken into account, including NLTE effects, size of the telescope
beam, and chemical abundances of the molecular species.  The NLTE level
populations have been calculate using a 2D Monte Carlo code (Hogerheijde \&
van der Tak \citeyear{hogervdtak:2000}). The reason for calculating the
full NLTE level populations instead of LTE is due to the low densities in
the upper layer of the disks (see van Zadelhoff et al. 2001). From these
populations, the emitted line profiles were constructed, convolving the disk
with a beam of 4.0\arcsec, the apparent size of the source on the sky at a
distance of 150 pc. In all cases the disk is assumed to be seen under an
inclination of 60\degr.  The abundance ratios with respect to H$_{\rm 2}$
are assumed to be constant ([CO]/[H$_{\rm 2}$]=8.$\times10^{-5}$) in the
entire disk. The abundances of the two isotopomers follow the isotopic
ratios of [$^{12}$C]/[$^{13}$C]=60 and [$^{12}$C]/[$^{18}$C]=500, similar to
the solar neighborhood.

The molecular line emission results are shown in Fig. \ref{fig-line}
and summarized in Table \ref{tab: hco}, in which the line emission
integrated over velocity is given. The three isotopomers, due to their
differences in abundance become optically thick at different heights in
the disk. CO reaches the $\tau$=1 plane close to the disk
surface, where temperature
and density are very similar in the two solutions. 
The line profiles shows only small changes between models
with differences
up to 5\% in the integrated line intensities. The $^{13}$CO becomes
optically thick in the intermediate layer showing slightly larger
differences between the two models (up to 15\%).  C$^{18}$O has a
optical depth of a few, and its lines form in the region of
the disk where the VEF and MEMO solutions differ more.
The differences between the two models are $\sim 30$\%. 
The CO isotopomers can be used to measure the vertical
temperature profile in disks (e.g.~Guilloteau \& Dutrey 
\citeyear{guilldutrey:1994}), and in this context
the differences  between different radiation transfer methods
become significant.

For each isotope, the difference between the more optically thick
3-2 and the less optically thick 6-5 follows the same pattern,
as seen in Fig.~\ref{fig-line}.
However, the difference in the ratio of integrated intensities
between models is small when
compared with the typical observational errors
(10-20 \%).

\begin{table}
\caption{\label{tab: hco} The integrated line emission for the CO,
$^{13}$CO and C$^{18}$O transitions.  All values have been derived after a
convolution with a beam of 4.00 \arcsec.}
\begin{tabular}{lllcllc}
\hline
transition & \multicolumn{3}{c}{VEF} & \multicolumn{3}{c}{MEMO} \\
           & \multicolumn{3}{c}{K km s$^{-1}$} & \multicolumn{3}{c}{K km s$^{-1}$} \\
           &  3-2 &  6-5 &  6-5/3-2 & 3-2 &  6-5 &  6-5/3-2 \\
\hline
CO         & 14.93 & 9.41 & 0.63  & 14.25 & 9.28 & 0.65 \\
$^{13}$CO  & 7.83  & 3.54 & 0.45  & 6.80  & 3.09 & 0.45 \\
C$^{18}$O  & 4.53  & 1.27 & 0.28  & 3.46  & 0.92 & 0.27 \\
\hline
\end{tabular}
\end{table}

 These calculations were performed assuming constant abundances throughout
the disk. In reality this will not be the case due to dissociation of
molecules in the upper layers by stellar and interstellar radiation and
freeze-out of molecules on the grain surface at $T < 20$K (Aikawa et
al.~\citeyear{aikawa:2002}). This latter process would enhance the difference
between the two models as more freeze-out is expected from the colder MEMO
model.

\section{Discussion}\label{sec-discussion}
The model described in this paper is relatively elementary in the sense that
all kinds of detailed microphysics are ignored. But within the limited
physics with which it is defined, and the geometrical simplifications made,
it is a reasonably accurate solution. The 1-D vertical radiative transfer
and the vertical pressure balance are solved without any approximations
other than the discretization itself. In this respect it is a step forwards
compared to the models of passive disks published in the literature so
far. The models worked out in this paper can be downloaded from a
website\footnote{{\tt http://www.mpa-garching.mpg.de/PUBLICATIONS/DATA/ 
radtrans/diskmodel/}}, which also features a couple of one-annulus test
setups for the radiative transfer problem in these disks.

The advantage of ignoring all the complex and detailed microphysics is that
we have a well-defined disk model with very few parameters. And all the
features of this model can be explained in terms of two pieces of physics
only: radiative transfer and hydrostatics. This simple model can then serve
as a basic model to which more physics can be added later, such as the
inclusion of dust settling and coagulation, the inclusion of dust
scattering, a treatment of internal heat processes such as viscous
dissipation, etc. A study of the effect of dust scattering on the
disk's structure and the outcoming spectrum is under way (Dullemond \& Natta
in prep.).

There are however a few uncertainties concerning the global structure of the
disk, even within the simple problem definition used in this paper. These
have to do with the possibility of self-shadowing. The present calculations
start from the {\em Ansatz} that the disk has a flaring shape, so that the
star's radiation can illuminate the disk at every radius. It then finds a
solution that is consistent with this Ansatz. But in reality perhaps part of
the disk might be non-flaring and reside in the shadow of the inner regions
of the disk. This could be the result of the history of formation of the
disk, or it may have to do with the outer parts of the disk becoming too
optically thin to sustain a positive derivative of $H_s/R$. But a
self-shadowed region in the disk could also be the end-product of an
instability (Dullemond \citeyear{dullemond:2000}; Chiang
\citeyear{chiang:2000}). In these cases the Ansatz of positive flaring index
is not confirmed, and the solutions might be different from the smooth
flaring disk solutions presented here. In order to investigate the
possibility of such alternative disk solutions, one must include full 2D
radiative transfer and full 2D hydrostatics into the model calculations. Due
to the enormous complexity of this problem, in particular at high optical
depths, we do not venture into this direction at present.

The models presented in this paper describe the middle and outer parts of a
passive flaring circumstellar disk. The very inner parts have a different
structure than the one described here. Very close to the star the dust has
evaporated and one is left with a pure gas disk.  If the gas is optically
thin, the inner rim of the dusty part of the disk is directly exposed to the
radiation of the star.  It will therefore be much hotter than the rest of
the disk, that is only irradiated on the surface. This inner rim will
therefore puff up and consistute a new component in the spectral energy
distribution of the disk (Natta et al.~\citeyear{nattaprusti:2001}). A
physical description of this inner rim, and the effects it has on the
complete disk structure is described by Dullemond, Dominik \& Natta
(\citeyear{duldomnat:2001}). But a full 2D model for this inner rim again
requires 2D radiative transfer, and is beyond the scope of this paper.

\section{Conclusion}
We have presented in this paper the results of calculations of the vertical
structure of irradiated circumstellar disks in hydrostatic equilibrium.  The
main difference with previous models (D'Alessio et al., Malbet et al. etc.)
is the full angle- and frequency-dependent treatment of radiative
transfer. It turns out that the frequency-dependent treatment of the
problem is of crucial importance for obtaining the correct vertical
structure (temperature and density) of the disk. 
Over most of the disk the midplane temperature is significantly lower than
the value predicted by a simpler radiative transfer method that uses 
Planck and Rosseland mean opacities. The
correct treatment of the angular dependence is of lesser importance. In
applications where computation time is critical, the use of the Eddington
approximation is reasonable, as long as the full frequency dependence of the
radiation field is taken into account.

If the disk model is used as input for further calculations, the need to
implement a reliable radiation transfer method is obvious. This is, for
example, the case when computing the expected intensity and profile of
molecular lines, as shown in \S 4 for the lines of the CO isotopomers, which
form at different depth in the disk.  As more realistic chemistry and line
transfer models are used, the differences between different disk models is
likely to become even more relevant (see van Zadelhoff and Dullemond, in
prep.), and the VEF models should be adopted to constrain physical and
chemical parameters in circumstellar disk environments.

The \SED{} is also affected by the change in vertical structure, but the
differences are not very large. In general the \SED{} compares reasonably
well with the results from both the \EAMO{} approach and the CG97/DDN01
model. This means that the model of Chiang \& Goldreich
(\citeyear{chianggold:1997}), with improvements described by Chiang et
al.~(\citeyear{chiangjoung:2001}) and Dullemond, Dominik \& Natta
(\citeyear{duldomnat:2001}), is a fairly robust model to compute the \SED{}
of T Tauri stars and Herbig Ae/Be stars.

\begin{acknowledgements}
We are grateful to E.~Kr\"ugel for assisting us with the testing of our
radiative transfer program and to M.~Hogerheijde and F.~van der Tak for use
of their 2D Monte Carlo code for line transfer. CPD acknowledges support
from the European Commission under TMR grant ERBFMRX-CT98-0195 (`Accretion
onto black holes, compact objects and prototars'). Astrochemistry in Leiden
is supported through a Spinoza grant from the Netherlands Organization for
Scientific Research (NWO).
\end{acknowledgements}

\appendix

\section{Radiative transfer with Variable Eddington Factors}
\label{sec-vet-method}
The idea at the basis of the variable Eddington factors method can be traced
back to the sixties (see e.g.~Mihalas \& Mihalas \citeyear{mihalmihal:1984}
and references therein).  For accretion disk theory the equations for this
method were presented by Hubeny (\citeyear{hubeny:1990}) and Malbet \&
Bertout (\citeyear{malbetbertout:1991}).  Here we present a slight variation
of this approach.  The advantage of our method is that it can be used up to
any optical depth even in cases where the opacity is non-grey and the
internal energy dissipation (and therefore the net flux) within the disk is
zero.

Define the mean intensity $J_\nu$, the Eddington flux $H_\nu$ and
the second moment of radiation $K_\nu$ as follows:
\begin{eqnarray}
J_\nu &=& \frac{1}{2}\int_{-1}^{1} I_{\mu,\nu} d\mu \\
H_\nu &=& \frac{1}{2}\int_{-1}^{1} I_{\mu,\nu}\,\mu\,d\mu
\label{eq-momdef-hnu} \\
K_\nu &=& \frac{1}{2}\int_{-1}^{1} I_{\mu,\nu}\,\mu^2\, d\mu 
\fullstop
\end{eqnarray}
By multiplying the transfer equation (\ref{eq-trans-de}) by resp.~$\mu^k$,
and integrating over $d\mu$ one obtains the $k$-th moment equation of
radiative transfer. The first two moment equations are:
\begin{eqnarray}
\frac{dH_\nu}{dz} &=& \rho \kappa_\nu ( B_\nu(T) - J_\nu) 
\label{eq-momeq-0}\\
\frac{dK_\nu}{dz} &=& - \rho \kappa_\nu H_\nu
\label{eq-momeq-1}
\fullstop
\end{eqnarray}
One can write the second moment $K_\nu$ as a dimensionless factor
times the mean intensity $J_\nu$:
\begin{equation}
K_\nu = f_\nu J_\nu
\comma
\end{equation}
where $f$ is the Eddington factor. For isotropic radiation the Eddington
factor is equal to $f=1/3$. For purely beamed radiation ($I_{\mu,\nu}=0$ for
$\mu\neq 1$) this factor is $f=1$. If one makes an assumption for this
value, then Eqs.(\ref{eq-momeq-0},\ref{eq-momeq-1}) form a closed set of
equations, which can be solved subject to boundary conditions. Often the
assumption is made that the radiation field will be more or less isotropic,
and therefore that $f_\nu=1/3$. This is the Eddington approximation, and
stands at the basis of many approximate transfer codes.

But if one had some way of {\em computing} (rather than guessing) the
Eddington factor, then the moment equations are not an approximation
anymore, and in fact will be exactly equivalent to the full transfer
equation. One way of doing so is to iteratively switch between the moment
equations and the real transfer equation. Start with a guess for $f_\nu(z)$
(take it for instance $1/3$), and solve the moment equations
(\ref{eq-momeq-0},\ref{eq-momeq-1}) to find the temperature. Then integrate
the formal transfer equation (\ref{eq-trans-de}) using the current
temperature. Using the resulting radiation field $I_{\mu,\nu}(z)$ one can
then compute the Eddington factor $f_\nu(z)$. Once this is done, one starts
all over again, using the newly computed Eddington factor. This process is
repeated until a converged solution for $T(z)$ is reached. Since in this
procedure the $f_\nu(z)$ is computed on the fly, one can be sure that this
solution is a real solution, and not an approximate one. It is important to
note that if this solution is inserted into the full set of transfer
equations (\ref{eq-trans-de},\ref{eq-therm-equil},\ref{eq-mean-int}),
then one will see that these are solved as well. In other words: the moment
equations were used to find a solution to the real transfer equations.

The advantage of the moment equations (\ref{eq-momeq-0},\ref{eq-momeq-1}) is
that they can be solved directly, using a two-point boundary value approach.
In fact, the frequency-integrated moment equations are even simpler to solve
directly. The frequency-integrated version of Eq.(\ref{eq-momeq-0}) is:
\begin{equation}\label{eq-momeq-freqint-h}
\frac{dH}{dz} = \rho \left(\kappa_P(T) \frac{\sigma}{\pi}T^4 
- \kappa_J J \right)
\comma
\end{equation}
where $\kappa_P(T)$ is the Planck mean opacity and
\begin{equation}\label{eq-momdef-h}
H=\int_0^\infty H_\nu d\nu
\fullstop
\end{equation}
Using the concept of energy conservation, one can replace the
right-hand-side of Eq.(\ref{eq-momeq-freqint-h}) with the source
term:
\begin{equation}\label{eq-mom-h-alt}
\frac{dH}{dz} = \frac{q}{4\pi}
\end{equation}
This equation can be trivially integrated from the equator (starting
with $H=0$) up to $z=z_{\max}$. At that point we can compute the value of
the frequency integrated mean intensity $J$:
\begin{equation}
J(z=z_{\max}) = H(z=z_{\max}) / \psi
\comma
\end{equation}
where
\begin{equation}
J=\int_0^\infty J_\nu d\nu
\comma
\end{equation}
and $\psi$ is the ratio of $H/J$ as computed from the full radiative
transfer. Using the frequency-integrated version of Eq.(\ref{eq-momeq-1}),
\begin{equation}\label{eq-integral-fj}
\frac{d(fJ)}{dz} = - \rho \int_0^\infty \kappa_\nu H_\nu d\nu
\comma
\end{equation}
(where $f$ is the $J$-mean of $f_\nu$), and starting from the value
of $J(z=z_{\max})$ computed above, one can integrate back towards the
equator to find $J(z)$. The temperature then follows from
\begin{equation}
\kappa_P(T) \frac{\sigma}{\pi} T^4 = \kappa_J J + \frac{q}{4\pi}
\comma
\end{equation}
where $\kappa_J$ is the $J_\nu$-mean of the opacity computed from the
full transfer. 

By iterating on the above procedure, one can quickly find solutions to the
transfer equation. An even quicker convergence can be reached by applying a
linear convergence amplifier like Ng's algorithm (Ng \citeyear{ng:1974}). 

Typically one needs about 70 gridpoints in $z$, 40 points in $\mu$ and,
dependent on the kind of opacity table and the wavelength range, about 60
points in $\nu$. It is important to make sure that there are enough $\mu$
gridpoints near $\mu=0$, because in plan-parallel geometry most of the
radiation is at small values of $\mu$. We use a logaritmic grid in $\mu$
spanned between $\mu_0\simeq 0.01$ and $1$.

It is important to note why, in Eq.(\ref{eq-integral-fj}), we did not use a
flux-weighted mean opacity $\kappa_H$ (or Rosseland opacity) multiplied by
the frequency-integrated Eddington flux $H$, as was suggested for instance
by Malbet \& Bertout (\citeyear{malbetbertout:1991}). The reason is that in
a disk without internal heat production, the total flux $H$ is zero. The
frequency-dependent flux $H_\nu$, on the other hand, is non-zero, consisting
of downward diffusive flux at short wavelengths and an equal amount of
upward flux at long wavelengths. For non-grey opacities one therefore has a
virtually zero $H$ close to the equator, but a very non-zero $\int_0^\infty
\kappa_\nu H_\nu d\nu$ at the same location. An evaluation of
$\kappa_H=\int_0^\infty \kappa_\nu H_\nu d\nu/\int_0^\infty H_\nu d\nu$
would therefore yield a diverging number, and the algorithm becomes
unstable, whereas the equation in the form of Eq.(\ref{eq-integral-fj})
yields a stable algorithm.

The method described in this section has proven to work well, and reaches a
solution to an accuracy of $10^{-4}$ in temperature within 11 iterations and
to $10^{-8}$ within 22 iterations, independent of the optical depth. On
a Pentium III with 850 MHz clock frequency a typical radiative transfer
problem is solved in about 5 seconds.

\subsection{Testing}
In order to verify the reliability of the code, a few tests are performed.
First of all we do internal consistency checks. Since the \VEF{} method
computes the $J$ and $H$ from both the full transfer as well as the moment
equations, one can a-posteriori check whether the two are the same. For all
our models this turns out to be indeed the case.
Then we compare the solution to the results from an independent \ALI{}
code. For those cases in which the \ALI{} in fact converges to a satisfiable
degree, we find that the solution are indeed very close to each other.

To further convince ourselves of the validity of the solutions provided by
the \VEF{} method, we applied the \VEF{} method to a problem with grey
opacities. For such a problem it is known that the \EAMO{} method should
work reasonably well. And as expected, 
the \VEF{} method is consistent with the \EAMO{} results.

\subsection{Convergence}
\begin{figure*}[ht!]
\centerline{
\includegraphics[width=6cm]{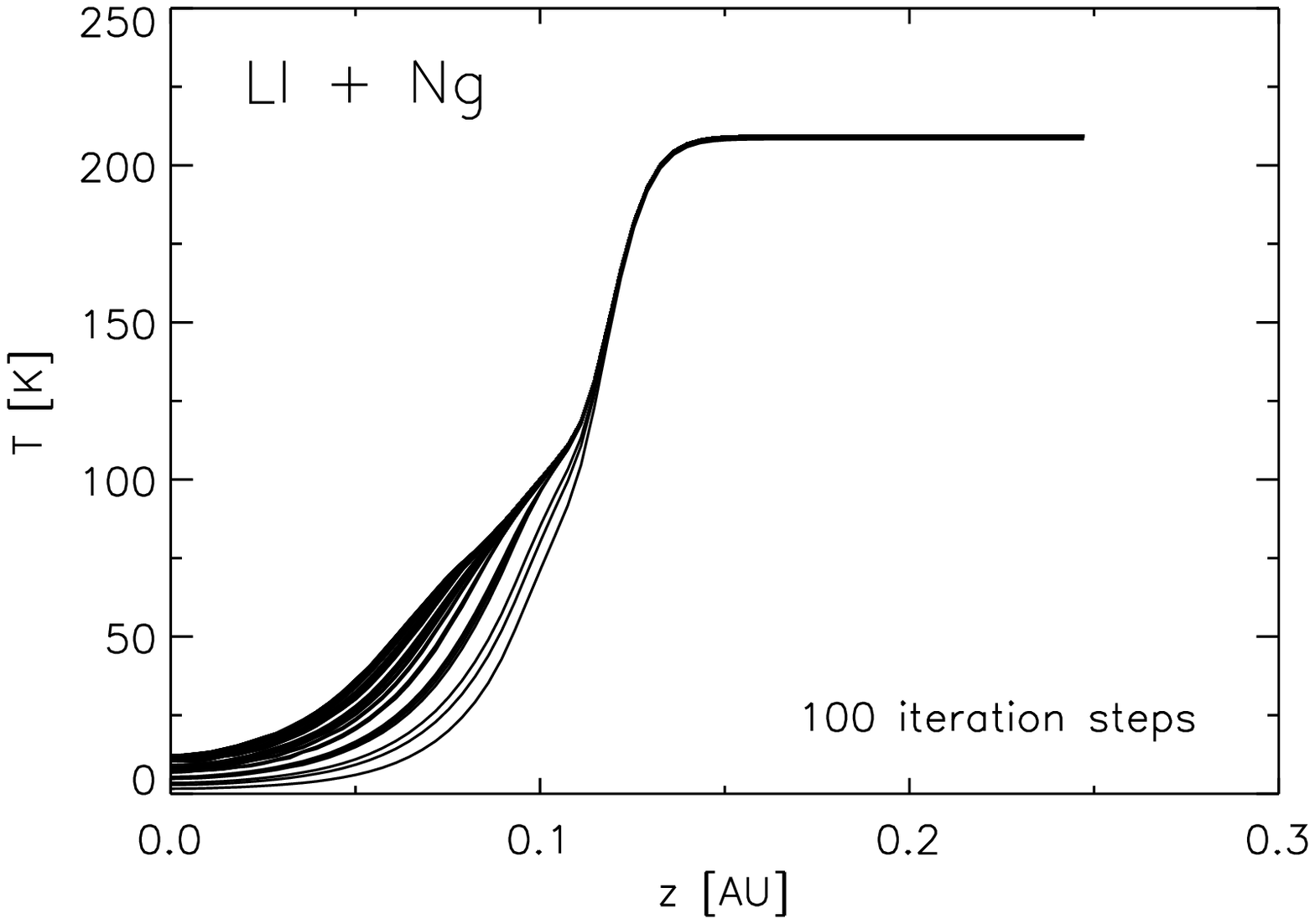}
\includegraphics[width=6cm]{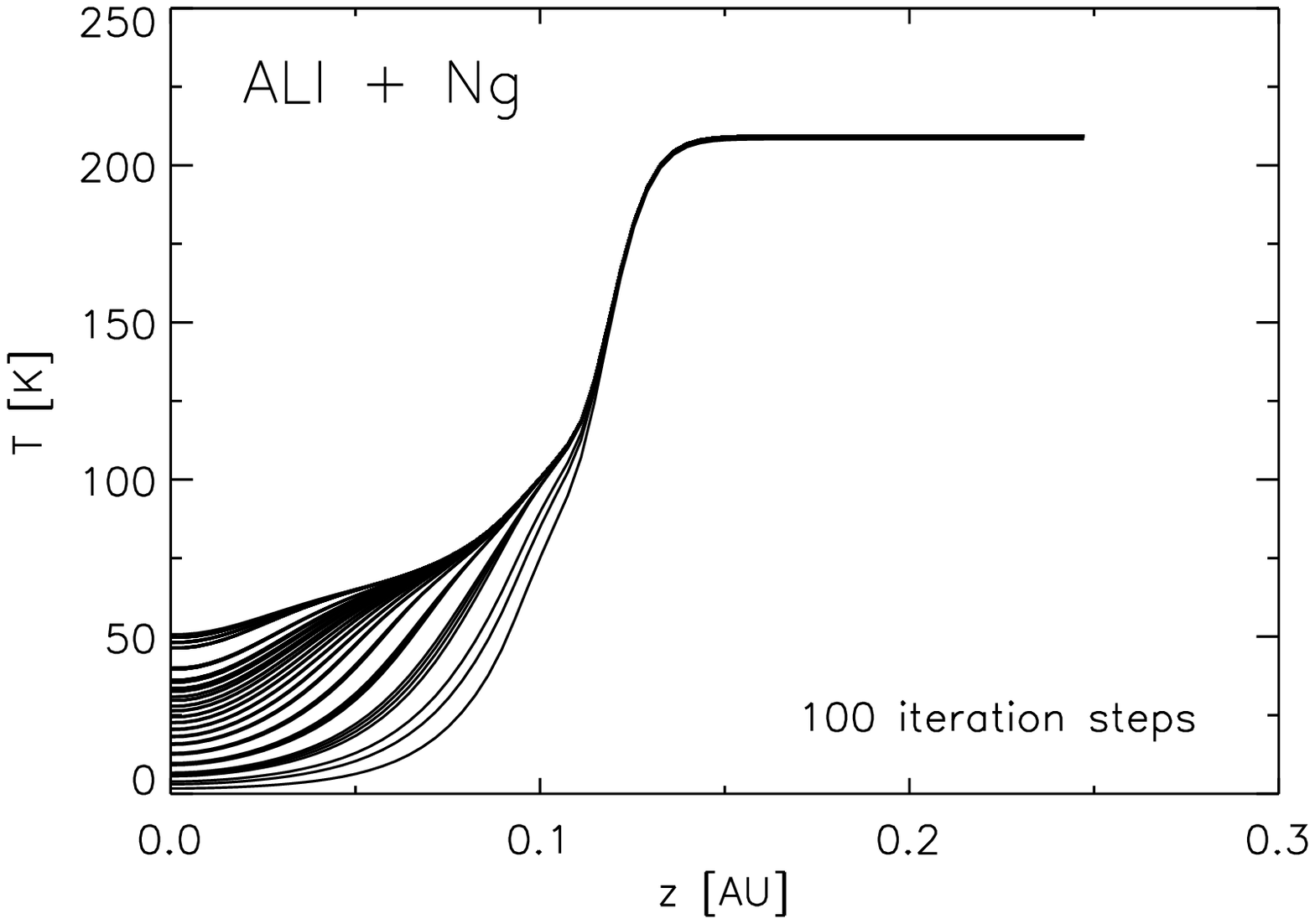}
\includegraphics[width=6cm]{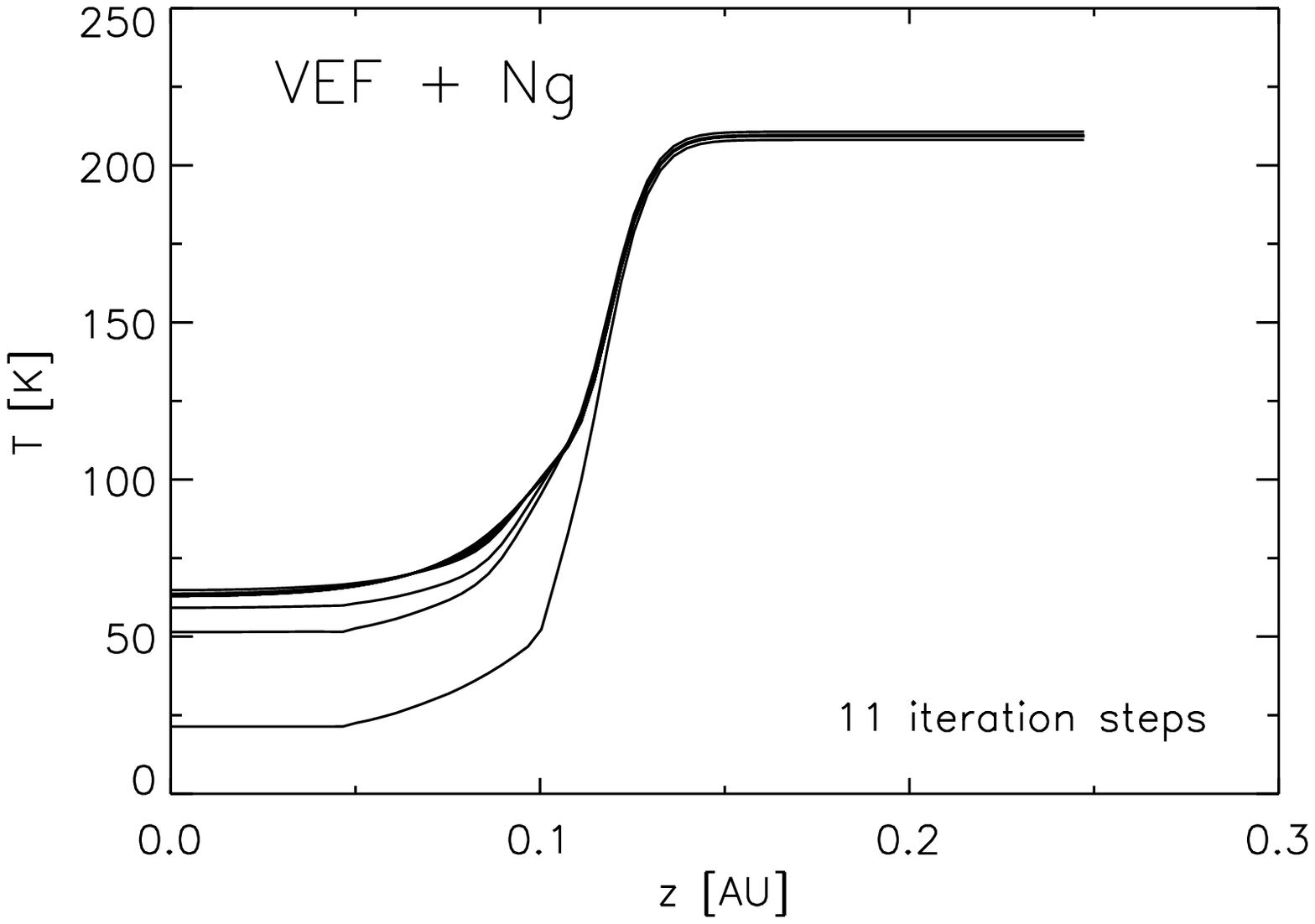}}
\caption{\label{fig-conv-hist} Comparison between different angle- and
frequency-dependent radiative transfer methods for the stage 2 transfer
problem. Plotted is the temperature at each iteration step. Left plot is for
Lambda Iteration (\LI), middle plot is for Accelerated Lambda Iteration
(\ALI), right plot is for Variable Eddington Factor (\VEF) method. All three
methods use Ng acceleration, which is responsible for the `jumps' in the
convergence.}
\end{figure*}
In order to demonstrate the advantage of the \VEF{} method over methods like
\LI{} and \ALI, we show here the convergence history for a test
problem. Consider a circumstellar disk which has $\Sigma=10^3\gram/\cm^2$ in
dust mass at 1 $\AU$ from a central star of $T_{\eff}=3000\kel$ and
$R_{*}=2.0\,R_{\odot}$. The vertical pressure scale height of the disk is
assumed to be $H_p=0.028\,\AU$, and the density profile is assumed to be
Gaussian. The flaring index equals $0.2$, and the opacity table used is for
astronomical silicate (Draine \& Lee \citeyear{drainelee:1984}). In
Fig.~\ref{fig-conv-hist} the convergence history for the temperature is
plotted using three different methods: \LI{} with Ng, \ALI{} with Ng and the
\VEF{} method with Ng. The convergence history is only shown for the first
100 iterations, within which only the \VEF{} method converges. After about
400 iterations, the \ALI{}+Ng method in fact also converges to within $1\%$
of the solution, but the \LI+Ng method does not converge even after 1000
iterations.

It should be noted that the method of Accelerated Lambda Iteration has not
been widely used for dust continuum transfer. It has been developed mainly
for line transfer, for which it has proven to be reasonably effective. We
adapted the algorithm for use with dust continuum transfer, and chose
as our approximate operator the diagonal of the full lambda operator.

\end{document}